\documentstyle[epsfig]{mn}
\def\msun{\rm M_{\odot}}
\def\etal{{et al.\ }}
\def\simlt{\mathrel{\rlap{\lower 3pt\hbox{$\sim$}}\raise 2.0pt\hbox{$<$}}}
\def\simgt{\mathrel{\rlap{\lower 3pt\hbox{$\sim$}} \raise 2.0pt\hbox{$>$}}}
\def\lsim{\mathrel{\rlap{\lower 3pt\hbox{$\sim$}}\raise 2.0pt\hbox{$<$}}}
\def\gsim{\mathrel{\rlap{\lower 3pt\hbox{$\sim$}} \raise 2.0pt\hbox{$>$}}}

\def\mbulge{M_{\rm Bulge}}
\def\msunpc3{\msun~{\rm {pc^{-3}}}}
\newcommand{\be}{\begin{equation}}
\newcommand{\ee}{\end{equation}}
\def\kms{{\rm\,km\,s^{-1}}}

\begin{document}

\title[Binary black holes]{Supermassive black hole binaries in gaseous and
stellar circumnuclear discs: orbital dynamics and gas accretion.}

\author[Dotti, Colpi, Haardt, Mayer]{M.~ Dotti$^1$, M.~Colpi$^2$, F.~Haardt$^1$, \& L. Mayer$^{3,4}$\\
$^1$ Dipartimento di Fisica e Matematica, Universit\`a  
dell'Insubria, Via Valleggio 11, 22100 Como, Italy\\
$^2$ Dipartimento di Fisica G.~Occhialini, Universit\`a degli Studi di Milano
Bicocca, Piazza della Scienza 3, 20126 Milano, Italy\\
$^3$  Institute for Theoretical Physics, University of Zurich, CH-8057, Zurich, 
Switzerland\\
$^4$ Institute of Astronomy, Department of Physics, ETH, Zurich, Wolfgang-Pauli Strasse,CH-8095 Zurich, Switzerland\\}

\maketitle \vspace {7cm}

\begin{abstract}
  The dynamics of two massive black holes in a rotationally supported
  nuclear disc of mass $M_{\rm disc} = 10^8 \msun$ is explored using
  N--Body/SPH simulations.  Gas and star particles are co--present in
  the disc. Described by a Mestel profile, the disc has a vertical
  support provided by turbulence of the gas, and by stellar velocity
  dispersion.  A primary black hole of mass $4\times 10^6 \msun$ is
  placed at the centre of the disc, while a secondary black hole is
  set initially on an eccentric co--rotating orbit in the disc plane.
  Its mass is in a 1 to 1, 1 to 4, and 1 to 10 ratio, relative to the
  primary.  With this choice, we mimic the dynamics of black hole
  pairs released in the nuclear region at the end of a gas--rich galaxy
  merger.  It is found that, under the action of dynamical friction,
  the two black holes form a close binary in $\sim 10$ Myrs.  The
  inspiral process is insensitive to the mass fraction in stars and
  gas present in the disc and is accompanied by the circularization of
  the orbit.
  We detail the gaseous mass profile bound to each black hole
  that can lead to the formation of two small Keplerian
  discs, weighing $\approx 2\%$ of the black hole mass, and of size $\sim
  0.01$ pc.  The mass of the tightly (loosely) bound particles
  increases (decreases) with time as the black holes spiral into
  closer and closer orbits. Double AGN activity is expected to occur
  on an estimated timescale of $\lsim 10$ Myrs, comparable to the
  inspiral timescale.  The double nuclear point--like sources that may
  appear during dynamical evolution will have typical
  separations of $\lsim 10$ pc.

\end{abstract}

\begin{keywords}
black hole physics -- hydrodynamics -- galaxies: starburst
-- galaxies: evolution -- galaxies: nuclei
\end{keywords}

\section{Introduction}

Collisions of spiral galaxies may lead to the formation of (ultra--)luminous
infrared galaxies (U--LIRGs). A clear
example is NGC 6240, a ULIRG composed of two
gas--rich galaxies on the verge of merging, hosting
two spatially distinct X--ray sources. 
Buried in a wide--spread star burst, these two hard X--ray sources 
have been identified as a pair of accreting massive black holes (MBHs) 
(Komossa et al. 2003; Risaliti et al. 2006). 

The majority of (U--)LIRGs are unsettled remnants of colliding galaxies 
undergoing episodes of intense star formation 
(see for a review Sanders \& Mirabel 1996). 
A large number of (U--)LIRGs hosts a central rotationally supported
massive gaseous disc (up to $10^{10}\msun$) extending on scales of
$\sim 100$ pc (Sanders \& Mirabel 1996; Scoville, Yun \& Bryant 1997; Downes
\& Solomon 1998; Tacconi et al. 1999; Bryant \& Scoville 1999; Greve
et al. 2006). 
As shown in numerical simulations, these discs may be the end--product of 
gas--dynamical and gravitational torques (excited during the merger), driving
large amounts of gas in the core of the remnant, where the MBHs are expected
to reside (Barnes \& Hernquist 1991, 1996; Kazantzidis et al. 2005). 
Inside these massive self--gravitating discs, MBH pairs continue
their dynamical evolution, and can accrete gas.

The formation and evolution of MBH pairs in merging galaxies have a
large number of potentially interesting astrophysical consequences.
In gas--rich mergers, MBH binaries can produce, beside double active nuclei, other peculiar 
features such as periodic modulations (as seen in OJ 287), and
wiggling jets (see Komossa 2006 for a review).
In dry mergers, binary MBHs can erode the inner stellar cusp (Milosavljevic et al. 2002), 
creating a population of hyper--velocity stars in galactic halos (Yu \& Tremaine 2003;
Brown et al. 2006). 
Ultimately, binary MBHs may eventually coalesce
emitting low--frequency gravitational waves
detectable by {\it LISA} (Bender et al. 1994; Hughes 2002; Sesana et
al. 2005). 

Mayer et al. (2006) have shown that in dissipative mergers,
large--scale inflows ending in the formation of massive turbulent
nuclear discs facilitate the pairing of the MBHs hosted in the parent
galaxies, and that an eccentric Keplerian binary forms. 
In this paper, we will 
explore the phase following the formation of the binary.
In a previous work (Dotti, Colpi \& Haardt 2006, hereafter DCH06), we 
studied the dynamical evolution of MBH pairs orbiting 
inside a massive  
gaseous disc (see also Escala et al. 2005), exploring different MBH mass ratios
and initial orbital eccentricities. We showed that, during MBH inspiral, 
gas--dynamical frictional forces 
circularize the orbits (if initially corotating with the disc),
and only after circularization a sizable amount of gas is captured by each MBH.  
The force resolution in such early simulations was
not sufficient to resolve the small accretion discs
that form around the inspiralling MBHs.

In the simulations presented in DCH06, the gaseous disc was embedded into a
larger scale, spherically symmetric stellar distribution. 
Fragmentation of the disc was prevented by using an adiabatic equation of
state, and star formation was, simply, neglected. Thus,
these simulations did not consider the possible co--presence of
an axisymmetric nuclear stellar disc. Indeed, nuclear
stellar discs are observed in many early and late type galaxies
(van den Bosch et al. 1998; Scorza \& van den Bosch 1998; Morelli et al. 2004;
Krajnovic \& Jaffe 2004; Lopes et al. 2006; Mueller Sanchez et al. 2006;
Davies et al. 2007). 
Nuclear stellar discs result from galaxy mergers and/or from
the secular evolution of bars (Krajnovic \& Jaffe 2004; Kormendy et al. 2005;
Ferrarese et al. 2006). 
Davies et al. (2007), observing young stellar discs in NGC 1068 and NGC 1097, 
estimate masses of $\sim 10^8 \msun$,
radii of $\sim 50$ pc, and a vertical scale heights of $\sim 5-10$ pc,
suggesting that the star velocity dispersion provides significant
pressure support.

Aiming at a better understanding of i) the role of a nuclear stellar disc
in driving the orbital evolution of the MBH binary, and ii) 
the formation of an accretion disc around each MBH, 
we run a new series of simulations employing the particle
splitting technique (Kitsionas \& Whitworth 2002), allowing us to improve 
the numerical resolution with respect to DCH06 and other 
similar simulations (Kazantzidis et al. 2005).
In the most accurate simulation, we can resolve, on sub--parsec
scales, the gravitational sphere of influence of
each MBH. For the first time, we detail the  
mass profile of the gas bound to the MBHs, so that we can 
conjecture on the accretion process during the MBH pairing and inspiral.

The paper is organized as follows. In Section 2 we describe
the equilibrium structure of the gaseous and stellar disc, and the initial
conditions for our different runs. In Section 3 we present the results
of our simulations carried on varying the MBH masses and the stellar
content in the disc, as well as the aforementioned 
simulation at higher resolution, and finally, in 
Section 4 we draw our conclusions.

\section{Simulation setup}

We follow the dynamics of MBH pairs in nuclear discs
using numerical simulations run with the N--Body/SPH code GADGET
(Springel, Yoshida \& White 2001). 

In our models, 
a primary MBH of mass $M_1$ is placed at the centre of
a gaseous and/or stellar disc, embedded in a larger scale
stellar spheroid. A secondary MBH 
of mass $M_2$ is placed on a bound orbit in the disc plane.

\begin{figure}
\begin{center}
\centerline{\psfig{figure=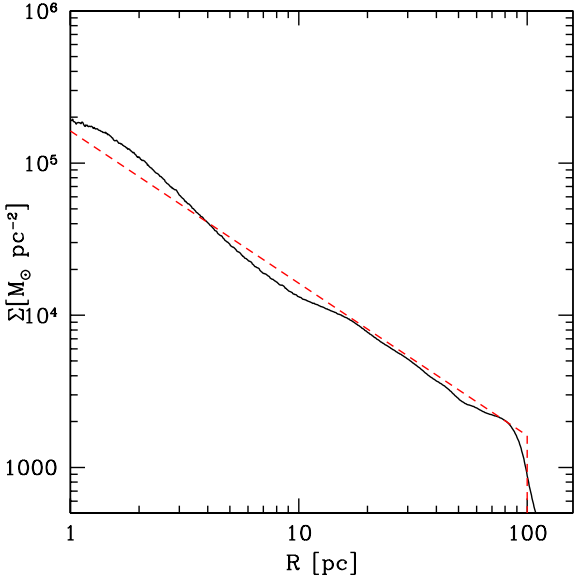,height=8cm}}
\caption{Gaseous disc surface density $\Sigma$ 
as a function of distance $R$ from the 
central MBH. Red dashed line refers to the
surface density of the Mestel disc (i.e., the disc
before it reaches equilibrium), while 
black solid line describes the equilibrium
profile after an elapsed time $\approx$ 3 Myrs.
}
\label{fig:001} 
\end{center}
\end{figure}

The gaseous disc is modeled with 235331 particles, has a total mass
$M_{\rm{Disc}}=10^8 \msun$, and follows a Mestel surface density profile  
\be 
\Sigma(R)=\Sigma_0\,\left({R_0\over R}\right), 
\ee 
where $R$ is the radial distance projected into the disc plane,
and $\Sigma_0$ is the density at scale radius $R_0$. 
Fig.~\ref{fig:001} shows the profile $\Sigma(R),$ in physical units.   
The disc is rotationally supported in $R$, has a finite radial extension
of 100 pc, and finite vertical thickness of 10 pc. Initially, the
gaseous particles are distributed uniformly along 
the vertical axis.
The SPH particles evolve adiabatically, and the
initial internal energy density profile scales
as:

\be
u(R)=K R^{-2/3},
\ee
where $K$ is a constant defined so that the Toomre parameter of the disc, 
\be \label{Toomre}
Q=\frac{\kappa c_{\rm s}}{\pi G \Sigma},
\ee
is $> 3$ everywhere, preventing disc fragmentation and
formation of  large scale over--densities, such as bars and 
spiral arms.
In equation~(\ref{Toomre}) $\kappa$ is the local epicyclic
frequency, and $c_{\rm s}$ the local
sound speed of the gas, that, for our choice of $K$, is $\approx 30 \kms$
at $R=50$ pc.
The internal energy of gas particles mimics the internal, unresolved
turbulence, and, as a consequence, $c_{\rm s}$ has to be considered as the
local turbulent velocity.

\begin{figure}
\begin{center}
\centerline{\psfig{figure=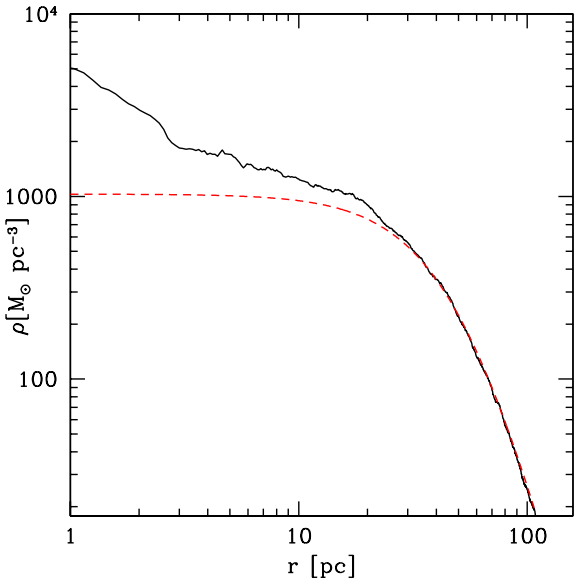,height=8cm}}
\caption{Density $\rho$ of the stellar spheroid
as a function of distance $r$ form the
central MBH. Red dashed line refers to the
initial Plummer profile, while black solid
line refers to the equilibrium profile.
}
\label{fig:002} 
\end{center}
\end{figure}

The spheroidal component (bulge) is modeled with $10^5$ collisionless
particles, initially distributed as a Plummer sphere with 
mass density profile (shown in Fig.~\ref{fig:002})
\be \rho (r)={3 \over 4 \pi}{\mbulge\over b^3}
\left(1+{r^2\over b^2}\right)^{-5/2}, \ee 
where $b$ $(=50$ pc $)$ is the 
core radius, $r$ the radial coordinate, and $\mbulge(=6.98
M_{\rm{Disc}})$ the total mass of the spheroid.  With such choice, the 
mass of the bulge within $100$ pc is 5 times the mass of the disc, as
suggested by Downes \& Solomon (1998).  

The primary MBH has a mass  $M_1=4\times 10^6\,\msun$,
while for the secondary we consider three different values, equal to
$4\times 10^5\,\msun$, $10^6\,\msun$ and $4\times 10^6\,\msun,$ respectively. 
The secondary MBH is placed inside the massive disc on an eccentric orbit 
with eccentricity $e\simeq 0.7$, at a separation of 50 pc from
the central MBH.
 
We evolve our initial composite model (bulge, disc and primary MBH)
for $\approx 3$ Myrs, until the bulge and the disc reach equilibrium. 
Given the initial homogeneous vertical structure of the disc, 
the gas initially collapses on
the disc plane exciting small waves that propagate through the system.
The spheroid is gravitationally stable on large scales ($r\gsim 20$ pc) and 
contracts in the central region until equilibrium is reached. 
Fig.~\ref{fig:001} and ~\ref{fig:002} show the 
equilibrium profiles, and in particular the density enhancement
in the central region of the spheroid. 
The stellar density increase does not affect the inner
region ($r\lsim 10$ pc) of the (denser) disc nor the dynamics
of the MBHs. 

We run four different sets of simulations (for a total of
12 according to the mass $M_2$ of the secondary MBH), assuming
a purely gaseous disc, and a disc in which 1/3, 2/3, and,
finally, all gas particles are turned into collisionless
particles, respectively. For each disc model with fixed star
fraction, we evolved the initial condition in isolation until
equilibrium is reached, as indicated above. The gas $\rightarrow$ star
conversion in the disc is obtained converting randomly selected
gas particles in collisionless particles. The latter own the
same mass and position of the initial gas particles, so that
the stellar disc component follows the same density distribution
of the gas. The velocities of the transformed particles are
modified by adding an isotropic component equal (in modulus)
to the local sound speed, preventing the vertical collapse
of the new disc. Gas particles are transformed at a constant
rate of $\approx \,160\, \msun {\rm yr}^{-1}$.  This rate prevents
spurious relaxation that could, in principle, change the
structure of the disc.  The outputs of this trial simulation
at three different times (t $\approx$ 0.2, 0.4, and 0.6 Myrs)
correspond to a fraction of 1/3, 2/3, and to the total disc
mass transformed in stars. These outputs are used as new
initial conditions for the three sets of simulations with an
axisymmetric stellar population.  We do not convert any gaseous
particle in stars when we follow the dynamics of the MBHs, so
that the disc stellar fraction remains constant. The secondary
MBH is inserted in the plane of the disc, after the composite
system has relaxed to equilibrium.

The number of particles used to model the two components allows for a
spatial resolution $\simeq 1$ pc, corresponding to the gravitational
softening which is the same for the collisionless, gaseous and MBH
particles. The thermodynamical parameters of the gaseous disc are
evaluated averaging over a subsample of 50 neighbours. 

The mass of the primary and the mass ratio $q= M_2 /M_1$ fix 
two characteristic scale--lengths. Following Merritt (2006), we define
the gravitational influence radius of the larger MBH ($r_{\rm inf}$)
as the radius of a sphere that encloses a mass (in gas and stars) equal to $2 M_1$. 
For separations smaller than $r_{\rm inf}$ the two MBHs form
a ``binary''. For the Plummer sphere and Mestel disc parameters used, 
$r_{\rm inf} \approx 6$ pc, so it is well resolved in our simulations. 
When the binary 
forms, we can estimate the hardening radius $a_{\rm h}$. 
Assuming a SIS profile,  
$a_{\rm h}=0.25 \,{q} \,r_{\rm inf}/(1-q)^2$ (Merritt 2006), and, for
$q=1$, $a_{\rm h}\approx 1$ pc, while for $q=0.25 \;(0.1)$ $a_{\rm h}
\approx 0.3 \;(0.1)$ pc.

The main input parameters of our simulations
are summarized in Table 1.  
We also performed a simulation at higher resolution, starting from an
intermediate snapshot of run A1, that will be detailed and discussed 
in section 3.2.

\begin{table}\label{tab:run}
\begin{center}
\caption{Run parameters}
\begin{tabular}{l@{   }c@{   }c@{   }c@{   }c@{   }c@{   }c@{   }}  
\hline
\\
run & $f_*^a$ & $M_{1}^b$ & $M_{2}^b$ & $M_{\rm Disc}^b$  & 
$M_{\rm Bulge}^b$ & $e$ \\
\\
\hline
\hline 
\\
A1$^{c}$ &   &    &  4   &     &     &  \\
A2& 0&  4 & 1    & 100 &  698 & 0.7\\
A3&   &  &  0.4   &     &     &  \\ 
\hline
B1&   &    & 4     &     &     &  \\ 
B2& 1/3& 4 & 1    & 100 & 698 & 0.7\\
B3&   &   & 0.4    &     &     &  \\
\hline
C1&   &    & 4   &     &     & \\
C2& 2/3& 4 & 1    & 100 & 698 & 0.7\\
C3&   &   & 0.4    &     &     &\\
\hline  
D1&   &   & 4    &     &     &\\
D2& 1& 4 & 1    & 100 & 698 & 0.7\\
D3&   &   & 0.4   &     &     &\\
\hline
\end{tabular}\\
\end{center}
\noindent
\footnotesize{$^{a}$ $f_*$: disc mass fraction in stars.}\\
\footnotesize{$^{b}$ Masses are in units of $10^6 \,\msun$.}\\
\footnotesize{~$^{c}$ Simulation A1 was run also using the  
particle splitting technique to improve force resolution
(down to $\simeq 0.1$ pc, see sect. 3.2).}\\
\end{table}

\section{RESULTS}

\subsection{Orbital decay and circularization}

\begin{figure}
\begin{center}
\centerline{\psfig{figure=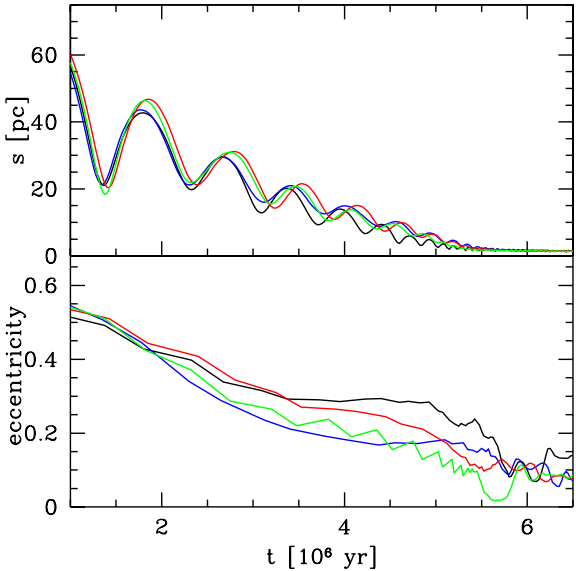,height=8cm}}
\caption{Equal mass MBHs. Upper panel: separations $s$ (pc) between the
MBHs as a function of time. Lower panel: eccentricity of the MBH binary as
a function of time. Black, blue, red and green lines refer to stellar
to total disc mass ratio of 0, 1/3, 2/3 and 1 (run A1, B1, C1, and D1) 
respectively.
}
\label{fig:003} 
\end{center}
\end{figure}

Fig.~\ref{fig:003} shows the MBH relative separation $s$ and
eccentricity as a function of time (upper and lower panel
respectively) for equal mass binaries (runs A1, B1, C1, and D1).
Regardless the fraction of star--to--gas disc particles, the
secondary MBH spirals in at the same pace,
reaching the force resolution limit of $\sim 1$ pc after 
$\sim 5$ Myrs. 
The velocity dispersion of stars is similar to the gas sound speed, and the 
two components share the same differential rotation. This is why  
the dynamical friction on $M_2$ caused by stars and gas is similar.  
As the orbit decays, the eccentricity decreases to
$e\lsim 0.2$.  This value is not a
physical lower limit, but rather 
a numerical artifact due to the finite resolution.
Circularization was found in DCH06 for a full gaseous disc. Here,
we show that circularization occurs regardless the nature of 
the disc particles (gas and/or stars). 
Note that circularization takes place well before the secondary 
feels the gravitational potential of $M_1$, so 
the MBH mass ratio does not play any role in the process. 

\begin{figure}
\begin{center}
\centerline{\psfig{figure=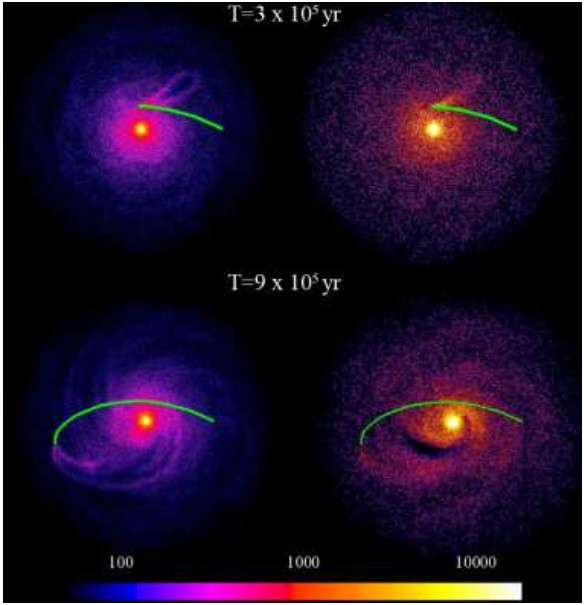,height=8cm}}
\caption{Snapshots from run C1. The panels show a face--on projection of
the disc and MBH positions at two different times. The color coding
indicates the z--averaged gas density (left panel) and star density
(right panel) in logarithmic scale (units: $\msun$/pc$^3$). The green
line traces the MBH counterclockwise prograde orbit. The gaseous (stellar)
disc radius is $R \approx 100$ pc.
}
\label{fig:004}
\end{center}
\end{figure}

To show how the circularization process works, consider run C1, 
where both stellar and gaseous particles are present in the disc. 
In Fig.~\ref{fig:004} we plot 
the gas (left panels) and
stellar  (right panels) densities 
in the disc at two different times, corresponding
to the first passage at pericentre (upper panels), and at 
apocentre (lower panels).  The green curve shows the counterclockwise 
corotating orbit of the secondary MBH.  Near to pericentre, the MBH has
a velocity larger than the local rotational velocity, so that
dynamical friction causes a reduction of the velocity
of the MBH. A wake of particles lags behind the MBH trail. On the
other hand, near to apocentre, the MBH velocity (mainly tangential) is
lower than the disc rotational velocity, and, in this case, 
the force increases the MBH angular momentum: the wake 
is dragged in front of the MBH trail.  
Thus, as a result of differential
rotation, the wake reverses its direction at apocentre 
accelerating tangentially
the MBH, leading to circularization of $M_2$ orbit (see also
Fig.~\ref{fig:003}). 
This seems a generic feature, regardless the disc
composition, as long as
the rotational velocity 
exceeds the gas sound speed and the stellar velocity
dispersion, as in all cases explored.
We remark that in spherical backgrounds, 
dynamical friction tends to increase the eccentricity, 
both in collisionless 
(Colpi, Mayer \& Governato 1999; van den Bosch \etal 1999; 
Arena \& Bertin 2007)
and in gaseous (Sanchez--Salcedo \& Brandenburg 2001) environments.

\begin{figure}
\begin{center}
\centerline{\psfig{figure=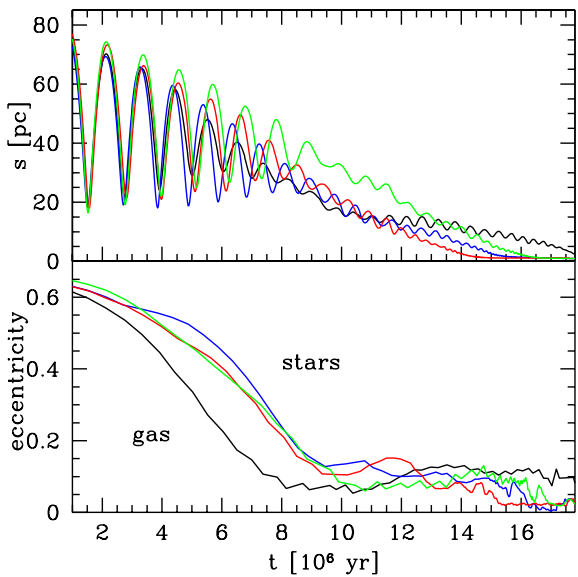,height=8cm}}
\caption{
  Same as Fig.~\ref{fig:003}, but for MBH mass ratio $1/4$ (runs ``2" in
  all sets).  Upper panel: separations $s$ (pc) between the MBHs as a
  function of time. Lower panel: eccentricity as a
  function of time. Black, blue, red and green lines refer to stellar
  to total disc mass ratio of 0, 1/3, 2/3 and 1 (run A2, B2, C2, and
  D2), respectively.  }
\label{fig:005}
\end{center}
\end{figure}

Fig.~\ref{fig:005} shows the MBH separation and eccentricity
evolution for the cases with $M_2=10^6\, \msun$ (runs ``2" for all sets). The
color coding is the same as in Fig.~\ref{fig:003}. As the dynamical 
friction time--scale is $\propto 1/M_2$ (for given disc properties), 
in this case $M_2$ orbital decay occurs on a time $\approx 4$ longer than in runs ``1". 
We can also observe that, in the case of a pure gaseous disc (run A2), 
circularization is more efficient compared to all other runs.
When the secondary MBH moves along the eccentric orbit, its motion is
supersonic for nearly all the
entire orbit. Since for a supersonic perturber in a gaseous background
dynamical friction 
is $\approx 2$ times more efficient that in a stellar environment
(Ostriker 1999), the circularization efficiency is increased (while, as discussed above, 
the orbital decay time--scale is nearly 
independent on the nature of the disc particles).

The same scalings hold for runs ``3", where $M_2=4\times 10^5\,\msun$. 
The differences in the orbital decay timescale 
among runs A3, B3, C3 and D3, 
are however larger than what found in runs ``1'' and ``2'', because of the smaller
secondary--to--background particles mass ratio. In runs ``3" a random component
of the motion of $M_2$ exists, mainly caused by the
stellar bulge. In fact, the mass ratio between $M_2$ and a single bulge
particle is 57, so that the background can not be
treated as a smooth fluid.

\subsection{Nuclear discs around the spiraling MBHs:
the high resolution run}

\begin{figure}
\begin{center}
\centerline{\psfig{figure=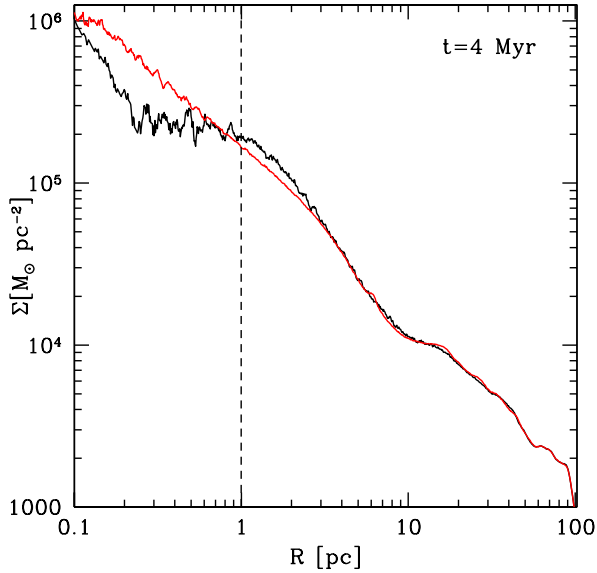,height=8cm}}
\caption{Surface density profile of the circum--nuclear gaseous disc in run
A1 at $t=4$ Myrs. Black line refers to the surface density in the low
resolution simulation, red line refers to the high resolution (split)
simulation. The dashed vertical line marks the resolution limit in the
un--split simulation.}
\label{fig:006}
\end{center}
\end{figure}

We run a higher resolution simulation to study the eccentricity and
orbital evolution on scales lower than 1 pc. The new
initial condition is obtained resampling the output of run A1 (for equal mass
MBHs) with
the technique of particle splitting. Resampling is performed when the
MBH separation is $\simeq 14$ pc (corresponding to $\simeq 4$ Myrs after
the start of the simulation). We split each gas particle into $N_{\rm ch}=10$ 
``child'' particles (Kitsionas \& Whitworth 2002), randomly
distributed around the position of the original parent particle within
a volume of size $\sim h_{\rm p}^3$, where $h_{\rm p}$ is the gravitational
softening (of the parent particles). The velocities and temperatures 
of the child particles are set equal to that of the parent ones. 
Each child particle has a mass equal to $1/N_{\rm ch}$
of the mass of the parent. Splitting is applied to all
particles whose distance from the binary centre of mass is $\leq 42$
pc, so that the total number of particles increases only by a
factor $\simeq 4$, while the local mass resolution in the split region
is comparable to that of a standard $\simeq 2 \times 10^6$ particle
simulation with uniform resolution. 
Our choice of the maximum 
distance for splitting is conservative, aimed at preventing that more massive,
unsplit gas particles reach the binary on a timescale shorter than
the entire simulation time. 
In the central split region, the high mass resolution achieved fulfills
the Bate \& Burkert (1997) criterion for gravitational softening values
down to $0.1$ pc. In other words, the resolution scales of hydrodynamical
forces (SPH smoothing) and gravitational forces (gravitational softening)
are similar. In particular, a sphere with radius $\approx 0.1$ pc
(corresponding to the new gravitational softening) centred on a
particle contains $\gsim 1$ SPH kernel ($N_{\rm neigh}=50$ particles for
our simulations).
In Fig.~\ref{fig:006} we compare the surface density profile of the
circum--nuclear gaseous disc in run A1 at $t=4$ Myrs, for the low and 
high resolution cases. The two profiles differ only below  
the scale of the low resolution limit $R\lsim 3$ pc. 
The decrease of the gravitational softening corresponds
to introducing a deeper potential well of the MBH
within a sphere defined by the former softening radius. 
Therefore, with the improved resolution,  the central
surface density increases as the gas reaches a new
hydrostatic equilibrium closer to the MBH, as shown
in Fig.~\ref{fig:006} (red line).
The lack of noticeable differences in the surface profile at
separations $R > 3$ pc confirms the accuracy of the particle splitting
technique.

\begin{figure}
\begin{center}
\centerline{\psfig{figure=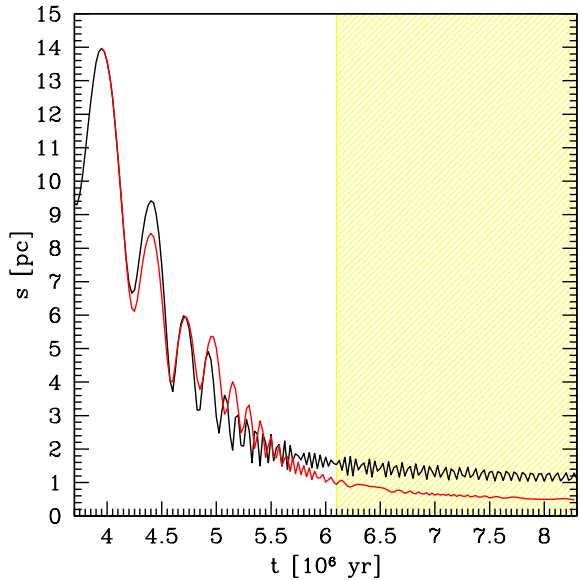,height=8cm}}
\caption{Separations $s$ (pc) between the MBHs as a function of time.
Red (black) line refers to the (un--)split run A1. The dashed area
highlight when the MBHs separation is $<1$ pc (corresponding to the 
low numerical resolution) in the split run A1.}
\label{fig:007}
\end{center}
\end{figure}

Results of the high resolution run are shown in Fig.~\ref{fig:007}.
The separation decays down to $0.1$ pc
in $\sim 10$ Myrs. We notice that the ellipsoidal torque 
regime, defined in Escala et al. 2004 (see also Escala et al. 2005)
and characterized by a fast orbital decay, is not present.
We attribute this difference to the different thermal state of the gas
surrounding the MBHs. For a comparison, we rescale the results of our
simulation to the parameters used in Escala et al. (2005). 
After rescaling, our initial disc is found to be 
hotter by a factor $\approx 2.4$ than
the hottest disc model presented in Escala et al. (2005). 
Since the efficiency of angular momentum
transport by ellipsoidal deformations decreases with
increasing temperature, in our hotter disc  
dynamical friction driven torques still
dominate over the ellipsoidal torques.

In the high resolution run, the dynamical
evolution of the MBHs is initially identical to the low resolution case, 
as shown in Fig.~\ref{fig:007}. 
Because of particle
splitting, the system granularity is reduced, 
and therefore the force resolution increases. 
In the high resolution run, the binary
decreases its eccentricity to $\approx 0$ (before
the new spatial resolution limit is reached).

\begin{figure}
\begin{center}
\centerline{\psfig{figure=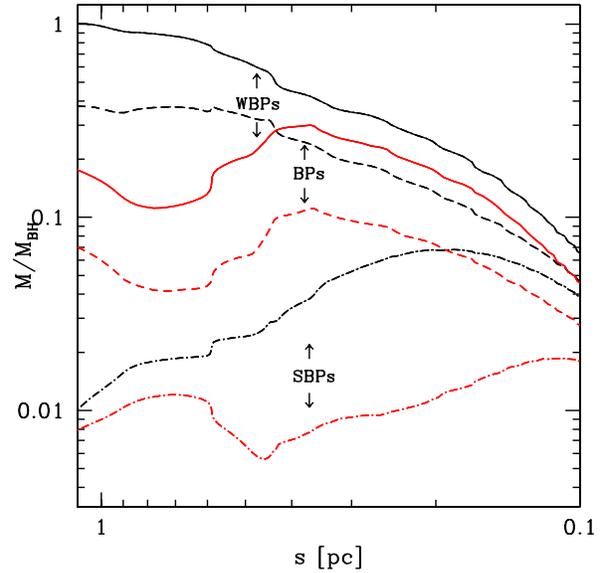,height=8cm}}
\caption{Mass of bound gas particles as a function of binary separation $s$
in the final stages (from  $\approx 6$ to $\approx 8.5$ Myrs) of the high
resolution simulation (split run A1).
Black lines refer to gas particles bound to the MBH initially at rest
(primary), red lines refer to gas particles bound to the secondary MBH.
Solid, dashed, and dot--dashed lines refer to WB, B, and SB particles.}
\label{fig:008}
\end{center}
\end{figure}

A resolution of 0.1 pc allows us to study the properties of the gas bound
to each MBH. 
To this purpose, it is useful to divide gaseous particles, bound to each MBH, 
into three subsets, according to their total energy
relative to the MBH. We then define weakly bound (WB), bound (B), and
strongly bound (SB) particles according to the following rule:
\be \label{bound}
E <
\left\{
\begin{array}{ll}
0  & (\mbox{WB}) \\
0.25\, W & (\mbox{B}) \\
0.5 \,W  & (\mbox{SB}),  
\end{array}
\right.  \ee 
where $E$ is 
the sum of the kinetic, internal and gravitational energy (per unit mass),
the latter referred to the gravitational potential $W$ of each 
individual MBH. 
Hereafter WBPs, BPs, and SBPs will denote particles satisfying the WB, B, or SB
condition, respectively. 
Note that, with the above definition, SBPs are a subset
of BPs, which in turn are a subset of WBPs.

We find that the mass collected by each 
MBH, relative to WB, B, and SB particles is
$M_{\rm WBP}\approx 0.85 M_{\rm MBH}\approx 3.4 \times 10^6 \msun$,
$M_{\rm BP}\approx 0.41 M_{\rm MBH}\approx 1.6 \times 10^6 \msun$, and 
$M_{\rm SBP}
\approx 0.02 M_{\rm MBH}\approx 8 \times 10^4 \msun$, respectively 
(here $M_{\rm MBH}=M_1=M_2$ of run A1).
These masses are of the same order for both the primary and secondary
MBH, and remain constant as long as the MBH separation is $s \gsim 1$ pc. 
At shorter separations WBPs and BPs are perturbed by the tidal field 
of each MBH, and at the end of the simulation $M_{\rm WBP}$ and 
$M_{\rm BP}$ are reduced as shown in Fig.\ref{fig:008}.  
During the same period of time, $M_{\rm SBP}$ associated to the 
primary (secondary) MBH increases
by a factor $\approx 4$ ($\approx 2.5$). 
This result is unaffected by numerical noise 
since the number of bound particles (associated to each class) 
is $\gsim 1$ SPH kernel ($N_{\rm neigh} = 50$).

\begin{figure}
\begin{center}
\centerline{\psfig{figure=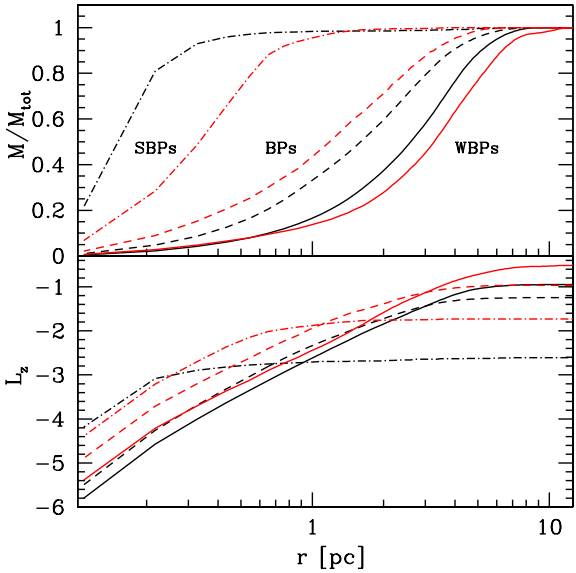,height=8cm}}
\caption{High--resolution simulation at $t=4.5$ Myrs.
Upper panel: gaseous cumulative mass fraction of bound
particles (for the SBPs, BPs, and WBPs, respectively)
as a function of the distance from each MBH.
Lower panel: $z$--component of cumulative orbital angular momentum, per
unit bound mass, as a function of the distance from each MBH.
Black (red) lines refer to the initial central primary 
(orbiting secondary) MBH.
Solid, dashed, and dotted--dashed lines refer to WBPs, BPs, and
SBPs, respectively.}
\label{fig:009}
\end{center}
\end{figure}

The radial density profiles of WBPs, BPs, and SBPs are well resolved
during the simulation.  The upper panel of
Fig.~\ref{fig:009} shows the normalized
mass within distance $r$ from each MBH, for WBPs,
BPs, and SBPs, at $t = 4.5$ Myrs, when the binary separation is
$\approx 5$ pc.
The lower panel of Fig.~\ref{fig:009} shows the specific angular
momentum perpendicular to the disc plane ($L_{\rm z}$) of the 
particles, computed in each MBH frame. Bound particles 
have a net angular momentum with respect to each MBH, and form a
pressure supported ellipsoid. 
The half--mass radius is similar for the two MBHs: 
$\simeq 3$ pc, $\simeq 1$ pc, and $\simeq 0.2$ pc for WBPs, BPs,
and SBPs, respectively.
The disc gas density can be as high as $10^7$ cm$^{-3}$. It is
then conceivable that, at these high densities, dissipative processes could be important, possibly reducing the gas internal (turbulent and
thermal) energy well below the values adopted in our simulations.
If any dissipative precess could reduce efficiently the gaseous
internal energy, we expect that the bound gas will form a 
cool disc with Keplerian angular momentum
comparable to what we found in our split simulation (see
Fig.~\ref{fig:009}).
Since $L_{\rm z}=\sqrt{G\,M_{\rm MBH}\, R_{\rm MBH,disc}}$, we obtain,
for the primary MBH, an
effective radius $R_{\rm MBH,disc} \approx 0.1$ pc ($0.03$ pc) for WBPs (BPs).
The secondary MBH is surrounded by particles with a
comparatively higher angular momentum (see red lines
in Fig.~\ref{fig:009}, lower
panel), with  a corresponding effective radius $R_{\rm MBH,disc}\approx 1$ 
(0.13) pc for WBPs (BPs). Finally, for SBPs, both holes have 
$R_{\rm MBH,disc} \ll 0.01$ pc, which is more than
an order of magnitude below our best resolution limit. These simple
considerations indicate that a more realistic treatment of gas
thermodynamics is necessary to study the details of gas accretion onto
the two MBHs during the formation of the binary, and the
subsequent orbital decay. Nonetheless, our simplified treatment
allows us to estimate a lower limit to the accretion timescale,
assuming Eddington limited accretion: 
\be t_{\rm
  acc}=\frac{\epsilon}{1-\epsilon}\,\tau_{\rm Edd} \,
\ln{\left(1+\frac{M_{\rm acc}}{M_{{\rm MBH,}0}}\right)}, 
\ee 
where $\epsilon$ is the radiative efficiency, $\tau_{\rm Edd}$ is the
Salpeter time, $M_{{\rm MBH,}0}$ is the initial MBH mass, and
$M_{\rm acc}$ is the accreted mass. 
Assuming $\epsilon=0.1$, Eddington limited accretion can last
for $\sim 30$ Myrs,
$\sim 15$ Myrs, and $\lsim1$ Myrs, if the MBHs accrete
all the WBPs, BPs, and SBPs, respectively.

\section{Discussion}

In this paper we have shown that dynamical friction against a
gaseous and/or stellar background is responsible for the inspiral of
MBH pairs inside massive nuclear discs.  A MBH binary forms at a
separation $\sim 5$ pc. The  
density and thermal distributions 
of the gas and star particles around 
the binary MBH 
are such that dynamical friction keeps acting on each MBH down to a
separation of $\approx 1$ pc since the coherence of the density wakes
excited by the MBHs is maintained down to this scale.  Thanks to the
particle splitting technique applied to the case of a purely gaseous
disc, we follow the binary orbital decay down to $\approx
0.1$ pc, and we show that angular momentum losses by friction, along
co--rotating orbits, reduce the orbital eccentricity to a value
consistent with zero.

Other processes neglected in our study can help to shrink the binary orbit. 
For example, three--body encounters with
background stars may become important at MBHs separations 
$\lsim r_{\rm inf} \approx 6$ pc
(Quinlan 1996; Milosavljevic \& Merritt 2001; Merritt 2006).
The cumulative effect of this collisional process,
studied mainly in spherical backgrounds, can lead to
an increase of the eccentricity, thus acting against the circularization
driven by the large--scale action of the gaseous and/or stellar disc
(Quinlan 1996; Aarseth 2003; Matsubayashi, Makino \& Ebisuzaki 2005;
Berczik et al. 2006; Sesana \etal in preparation). The effect of three body
encounters on the MBH orbits when stars form a rotationally supported disc
is still unclear.
 
We also quantified the structure of gas particles
which become bound to each MBH during the orbital evolution. 
We found that during the orbital inspiral, a gas mass $\simeq 50 \%$ of the MBH mass is conveyed 
inside the MBH sphere of influence, and that a gas mass $\approx 2\%$ of 
the MBH mass binds deeply to each single MBH.
The radial distribution of the most  
bound gas particles suggests that an active Eddington--limited
accretion phase may set in, for a time  $\lsim 1$ Myrs around both MBHs.
Since we have neglected gas cooling in the simulation, this 
mass likely represents a lower limit. 
The active phase could last for a longer time ($\gsim 10$
Myrs), comparable to the inspiral timescale, 
if all the bound mass is accreted.
This highlights the possibility of revealing
double AGN activity, on spatial scales $s\lsim 10$ pc. However, since we expect
that star formation is still ongoing in the disc while the MBHs spiral in, 
double AGN activity could be heavily obscured.
In the same high resolution simulation, we measured the cumulative
angular momentum 
(perpendicular to the disc) of particles bound to each
individual MBH, allowing an estimate of 
the size of the Keplerian disc that
is expected to form. We found disc sizes of $\approx 0.1$ pc and $\ll 0.01$ pc for the bound and the strongly bound gas components, respectively.
Such small extension of the accretion disc around each MBH
could preserve the gas against 
tidal perturbation and stripping, at least until the binary reaches separations
of the same order.

We plan to carry on higher resolution simulations with
more realistic input physics and radiative cooling,  
in order to follow the last phases of the gas--dynamical
evolution of the nuclear discs. In particular, 
we aim at tracing the expected transition of the two discs into a 
single, circumbinary disc surrounding the two MBHs.

Armitage \& Natarajan (2002) have shown that the interaction
between the two MBHs and the circumbinary disc 
can drive the binary to 
coalescence in $\sim 10$ Myrs, for MBH separations $\lsim 0.1$ pc.
They have also shown that, during the process, 
the MBH binary orbital eccentricity increases (Armitage \& Natarajan 2005).
Our planned future simulations will test in a self--consistent way such predictions, 
and will explore the possibility of exciting 
flaring activity during the latest phases
of MBH binary inspiral. This will constrain
the properties of an electromagnetic precursor associated to 
the binary coalescence (Armitage \& Natarajan 2002; Kocsis et al. 2005;
Milosavljevic \& Phinney 2005; Dotti \etal 2006).

\section*{Acknowledgments}

\end{document}